\documentclass[conference]{IEEEtran}
\IEEEoverridecommandlockouts
\usepackage{cite}
\usepackage{amsmath,amssymb,amsfonts,amsthm}
\usepackage{algorithm}
\usepackage{algorithmic}
\usepackage{bm}
\usepackage{multirow}
\usepackage{subcaption}
\usepackage{framed}
\newtheorem{proposition}{Proposition}
\usepackage{graphicx}
\usepackage{dsfont}
\usepackage{textcomp}
\usepackage{balance}
\usepackage{xcolor}
\usepackage{ativ}
\usepackage{mathtools}
\usepackage{tikz}
\def\<#1>{\mathinner{\langle#1\rangle}}
\usetikzlibrary{automata, arrows.meta, positioning}
\theoremstyle{definition}
\allowdisplaybreaks

\newtheorem{definition}{Definition}
\newcommand{\inner}[1]{\left\langle#1\right\rangle}

\def\BibTeX{{\rm B\kern-.05em{\sc i\kern-.025em b}\kern-.08em
    T\kern-.1667em\lower.7ex\hbox{E}\kern-.125emX}}

\begin{document}
\title{Universal Caching}

\author{\IEEEauthorblockN{Ativ Joshi}
\IEEEauthorblockA{School of Technology and Computer Science\\
Tata Institute of Fundamental Research\\
Mumbai 400~005, India \\
Email: ativ@cmi.ac.in}

\and

\IEEEauthorblockN{Abhishek Sinha}
\IEEEauthorblockA{School of Technology and Computer Science \\
Tata Institute of Fundamental Research\\
Mumbai 400~005, India \\
Email: abhishek.sinha@tifr.res.in}
}
\maketitle

\begin{abstract}
    In learning theory, the performance of an online policy is commonly measured in terms of the \emph{static regret} metric, which compares the cumulative loss of an online policy to that of an optimal benchmark in hindsight. In the definition of static regret, the action of the benchmark policy remains \emph{fixed} throughout the time horizon. Naturally, the resulting regret bounds become loose in non-stationary settings where fixed actions often suffer from poor performance.
    In this paper, we investigate a stronger notion of regret minimization in the context of online caching. In particular, we allow the action of the  benchmark at any round to be decided by a finite state machine containing any number of states. Popular caching policies, such as \textsc{LRU} and \textsc{FIFO}, belong to this class. Using ideas from the universal prediction literature in information theory, we propose an efficient online caching policy with a sub-linear regret bound. To the best of our knowledge, this is the first data-dependent regret bound known for the caching problem in the universal setting. We establish this result by combining a recently-proposed online caching policy with an incremental parsing algorithm, namely Lempel-Ziv '78. Our methods also yield a simpler learning-theoretic proof of the improved regret bound as opposed to the involved problem-specific combinatorial arguments used in the earlier works.      
    \end{abstract}

\section{Introduction and Related Work} \label{intro}
\begin{figure}
\centering
\includegraphics[scale=0.35]{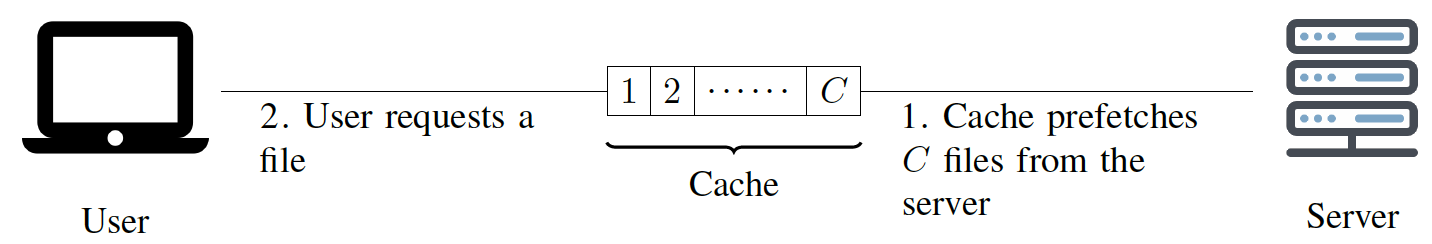}
\caption{\small{Setup for the caching problem}}
\label{caching}	
\end{figure}

\IEEEPARstart{W}{e} investigate the standard caching problem from an online learning perspective \cite{krishnan1998optimal, SIGMETRICS20, paschos2019learning, paria2021, mukhopadhyay2021online}. Consider a library consisting of $N$ unit-sized files $\{1,2,\ldots, N\} \equiv [N],$ and a cache of storage capacity $C$ (typically $C \ll N$). The system evolves in discrete rounds. At the beginning of round $t$, an online caching policy $\pi$ prefetches (possibly in a randomized fashion) a set of $C$ files, denoted by the incidence vector $\bm{y}_t \in \{0,1\}^N,$ where $||\bm{y}_t||_1=C$. After that, the user requests a file, which is denoted by the incidence vector $\bm{x}_t \in \{0,1\}^N,$ such that $||\bm{x}_t||_1=1$ (see Figure \ref{caching})\footnote{We will be using the (one-hot encoded) vectorized symbols $\bm{x}_t \in \{0,1\}^N$ and the corresponding scalars $x_t \in [N] $ interchangeably throughout the paper. }. The file request sequence $\{x_t\}_{t \geq 1}$ could be adversarial. In the case of a \emph{cache-hit}, which occurs when the requested file is present in the cache, the policy receives a unit reward. In the complementary event of a \emph{cache-miss}, the policy receives zero rewards. For simplicity, we do not charge any cost for file downloads (see \cite{mukhopadhyay2021online} for a model with download cost). Thus, the reward accrued by the policy at round $t$ is given by $\langle \bm{x}_t, \bm{y}_t \rangle $. The goal of the caching policy is to achieve a hit rate close to that of an optimal offline finite-state prefetcher ($\textsc{FSP}$) described next.

\begin{definition}[Finite State Prefetcher (FSP) \cite{krishnan1998optimal}]
An FSP is described by a quintuple $(\mathcal{S}, [N], g, f, s_0),$ where $\mathcal{S}$ is a finite set of states, $[N]$ is a library of $N$ files,  $g: \mathcal{S} \times [N] \to \mathcal{S}$ is the state transition function, $f: \mathcal{S} \to [N]^C$ is a possibly randomized prefetching policy that caches a set of $C$ files depending on the current state, and $s_0$ is the initial state. The components of an FSP without the prefetcher function $f$ is known as a Finite State Machine (FSM).  
\end{definition}

%
%
%
%
%
Let $x_1, x_2, \ldots $ be an $N$-ary sequence denoting the file requests. On round $t$, an FSP $\hat{\pi},$ which is currently at state $s_t$, first prefetches (possibly randomly) a set of $C$ files given by $\bm{\hat{y}}_t(\hat{\pi})=f(s_t)$, observes the file request $x_{t}$ for round $t$, incurs cache hits/misses, and then finally changes its state to 
$s_{t+1} = g(s_t, x_t).$ The reward obtained by an FSP $\hat{\pi}$ at round $t$ is given by the inner-product $\langle \bm{x}_t,  \bm{\hat{y}}_t \rangle.$ Denote the set of all FSPs containing at most $s$ states by $\mathcal{G}_s$ and define the set of all FSPs by $\mathcal{G} = \cup_{s=1}^\infty \mathcal{G}_s$\footnote{To be precise, the class $\mathcal{G}$ is parameterized by the numbers $N$ and $C$. Since these parameters will be clear from the context, we drop the parameters to avoid cluttering the notations.}. Informally, our objective is to design an online caching policy $\pi$ that performs as well as the best FSP in hindsight that knows the entire file request sequence a priori. Quantitatively, our goal is to design an online caching policy $\pi$ that attains a sublinear bound uniformly for all file request sequences for the regret metric $\mathcal{R}^{\pi}_T$ defined below:
\begin{eqnarray} \label{regret-def}
	\mathcal{R}_T^\pi = \sup_{\{\bm{x}_t\}_{t\geq 1}}\bigg(\max_{\hat{\pi} \in \mathcal{G}}\sum_{t=1}^T  \langle \bm{x}_t, \hat{\bm{y}}_t (\hat{\pi}) \rangle - \sum_{t=1}^T  \langle \bm{x}_t, {\bm{y}}_t (\pi)\rangle\bigg).
\end{eqnarray}

%
%
%
%
%
\begin{figure}[t]
    \begin{subfigure}{\columnwidth}
    \begin{tikzpicture}[node distance = 3cm, on grid, auto]
        \node (s0) [state, initial, initial text={}] {$s_0$};
        \node (s1) [state, right=of s0] {$s_1$};
        \node (s2) [state, right=of s1] {$s_2$};
    
        \path [-stealth, thick]
            (s0) edge node {\texttt{2}} (s1)
            (s1) edge node {\texttt{1,3}} (s2)
            (s2) edge [bend right] node[above] {*} (s0)
            (s0) edge [loop below] node {$*\backslash \{\texttt{2}\}$} ()
            (s1) edge [bend left] node {\texttt{2,4,5}} (s0);
    \end{tikzpicture}
    \caption{State transition function $g(s,x)$ of the FSM.}
    \end{subfigure}
    \hspace{0.4cm}
    
    \begin{subfigure}{0.40\columnwidth}
    \centering
    \begin{tabular}{l|l|l|l|l|l|l|}
        \multicolumn{2}{c}{} & \multicolumn{5}{c}{\small{Symbol} ($x$)}\\
        \cline{2-7}
        && \bfseries \texttt{1} & \bfseries \texttt{2} & \bfseries \texttt{3} & \bfseries \texttt{4} & \bfseries \texttt{5}\\ \cline{2-7}
        \multirow{3}*{\rotatebox{90}{States ($s$)}}
        &$\bm{s_0}$ & 0 & 4& 0& 0& 1\\ \cline{2-7}
        &$\bm{s_1}$ & 1 & 0& 2& 1& 0\\ \cline{2-7}
        &$\bm{s_2}$ & 0 & 0& 0& 1& 2\\ \cline{2-7}
    \end{tabular}
    \caption{Frequency $N_T(s,x)$.}
    \end{subfigure}\hfill
    \begin{subfigure}{0.40\columnwidth}
        \centering
        \begin{tabular}{|l|l|}
            \multicolumn{1}{c}{} & \multicolumn{1}{c}{$f^*(s)$}\\
            \hline
            \textbf{$s_0$} & \multicolumn{1}{l|}{\{2,5\}} \\ \hline
            \textbf{$s_1$} & \multicolumn{1}{l|}{\{1,3\}} \\ \hline
            \textbf{$s_2$} & \{4,5\} \\ \hline
            \end{tabular}
        \caption{Optimal prefetcher function $f^*(s)$}
        \end{subfigure}
    \caption{\small{This figure illustrates the computation of the optimal offline prefetcher for a given $3$-state FSM for the $5$-ary input sequence of length $T=12$ given by $(2,1,5,2,3,5,2,4,5,2,3,4).$ We assume that the size of the cache is $C=2$. The state transition function $g(\cdot)$ is shown in part (a) of the figure. The variable $N_T(s, x),x\in[5],$ denotes the number of times the file $x$ was requested while the FSM was visiting state $s$. For the given request sequence, the sequence of states visited by the FSM is given by $(s_0,s_1,s_2,s_0,s_1,s_2,s_0,s_1,s_0,s_0,s_1,s_2).$ Upon counting the frequency of the requests at each state, it is easy to see that the optimal prefetcher for the given FSM is $f^*(s_0)=\{2,5\}$, $f^*(s_1)=\{1,3\}$ and $f^*(s_2)=\{4,5\}$. The fraction of cache misses conceded by the optimal offline FSP is $0.083$.}}
    \label{fig:fsm}
\end{figure}
A brief discussion on the benchmark class $\mathcal{G}$ used in the performance metric \eqref{regret-def} is in order. In the case of the standard static regret minimization problems, the action of the offline benchmark $\hat{\bm{y}}_t$ remains constant throughout the entire time horizon of interest \cite{cesa2006prediction}. A number of recent papers studied the static regret minimization problem in the context of caching and proposed efficient online policies achieving sublinear regret \cite{SIGMETRICS20, mukhopadhyay2021online, paschos2019learning, paria2021, paschos2020cache, li2021online}. However, in terms of the absolute performance (total number of cache hits), these policies may perform poorly in ``non-stationary'' settings where the best offline static cache configuration has a poor hit rate. In the online learning literature, several generalizations of the  static regret metric have been proposed to quantify the performance of policies in non-stationary environments. For example, the \emph{Tracking Regret} metric allows changing the benchmark a fixed number of times within a given time horizon \cite{herbster1998tracking, cesa2012mirror, chen2020minimax}. The \emph{Adaptive Regret} metric compares the performance of an online policy over any arbitrary sub-interval $[s,e]\subseteq[T]$ with the best static policy $\pi^*[s,e]$ for that sub-interval \cite{hazan2007adaptive,daniely2015strongly, zhang2018dynamic, adamskiy2012closer}. In \emph{Dynamic Regret}, the benchmarks are allowed to vary slowly with time, subject to certain regularity constraints \cite{besbes2015non, zhang2018dynamic, jadbabaie2015online}. The FSP benchmark considered in the paper includes a rich class of comparators, which arises naturally in many contexts. In Section \ref{fsp-examples} of the Appendix, we show that popular caching policies with optimal competitive ratios, such as \textsc{LRU} and \textsc{FIFO}, belong to this class.


The seminal paper \cite{feder1992universal} considers a special case of the regret minimization problem \eqref{regret-def}, which, in our setup, corresponds to a library of size $N=2$ and a cache of capacity $C=1.$ In this context, the authors proposed an efficient universal prefetching policy by utilizing the Lempel-Ziv incremental parser \cite{ziv1978compression}. Follow this up, the paper \cite{krishnan1998optimal} considered the online caching problem with arbitrary values for $N$ and $C,$ and proposed a  universal caching policy achieving a sublinear regret. One of the key contributions of \cite{krishnan1998optimal} is the design of a new prefetching policy that is competitive against a single state FSP. In this paper, we give a tighter \emph{data-dependent} regret bound by utilizing a recent online learning policy obtained by combining the standard \textsc{Hedge} algorithm with Madow's sampling \cite{k-experts, cesa2006prediction, madow1949theory, tille2006sampling}. These improved bounds are obtained by using general learning-theoretic techniques, as opposed to the involved combinatorial arguments employed in \cite{krishnan1998optimal}. We also mention the paper \cite{pandurangan2010universal} which proposes a universal caching policy that is constant-factor optimal in the stochastic setting.


\section{Characterization of Finite-State Prefetchers}\label{sec:headings}
Before designing online policies, we first characterize the offline performance of the FSPs. In particular, we show that with almost no loss of generality, our attention can be restricted to a sub-class of FSPs, known as \emph{Markov Prefetchers}. 
\paragraph*{Characterization of the Optimal Offline Prefetcher}
Assume that an FSM $\mathcal{M}$ is run with the file request sequence $x_1^T.$  In this case, the optimal offline prefetcher $f^*,$ that maximizes the cumulative hits, is easy to determine (see Figure \ref{fig:fsm} for an illustration). Let the variable $N_T(s,i)$ denote the number of times the $i$\textsuperscript{th} file was requested while the FSM was visiting the state $s$. Let the set $\mathcal{A}_s$ denotes the most frequently-requested collection of $C$ files while the FSM was on the state $s$. Since the set of the prefetched file depends only on the current state of the FSP, the optimal offline prefetcher function for $\mathcal{M}$ is given by $f^*(s)= \mathcal{A}_s.$ We now recall a special sub-class of FSPs, known as \emph{Markov Prefetchers}, that plays a central role in universal caching.

\begin{definition}[$k$\textsuperscript{th}-order Markov Prefetcher \cite{feder1992universal}]
	A $k$\textsuperscript{th} order Markov Prefetcher is a sub-class of FSPs with $N^k$ states, where the state at round $t$ is given by the $k$-tuple of the previous $k$ file requests, \emph{i.e.,} $s_t=(x_{t-1}, x_{t-2}, \ldots, x_{t-k}).$ The state transition function $g(\cdot)$ is defined naturally using a shift operator.  
\end{definition}
For any given file request sequence $x^T_1,$ let  $\tilde{\pi}_S(x^T_1)$ denote the maximum fraction of cache hits \footnote{For notational conveniences, we work with cache hits rather than cache misses as in \cite{feder1992universal}. We use the tilde symbol on the top of the variables to emphasize that they represent offline quantities.} achieved by any FSP containing at most $S$ states and $\tilde{\mu}_k(x^T_1)$ denote the maximum fraction of cache hits achieved by a $k$\textsuperscript{th} order Markov prefetcher.  The following result, which is a generalization of \cite[Theorem 2]{feder1992universal} shows that the Markov Prefetchers are asymptotically optimal in the class of FSPs. 

\begin{theorem}\label{thm:FSP-MP1}
  The hit rate of a Markovian prefetcher of a  sufficiently large order $k$ exceeds the hit rate of any FSP with a fixed ($S$) number of states (up to a vanishingly small term). In particular, for any file request sequence $x^T,$ we have
    \begin{equation}
\pitld_{S}(x^T)-    \mutld_k(x^T)  \leq \min \bigg(1-C/N, \sqrt{ \frac{{\ln S}}{2(k+1)}}\bigg).
    \end{equation}
\end{theorem}
Please refer to Section \ref{proof:fsp-mp} of the Appendix for the proof of Theorem \ref{thm:FSP-MP1}. The proof closely follows the arguments for the binary case given in \cite{feder1992universal}. The message conveyed by Theorem \ref{thm:FSP-MP1} is that, in order to be competitive with \emph{any} FSM with a finite number of states $S$, an online policy only needs to be competitive with respect to a Markov prefetcher of a sufficiently large order ($k \gg \ln(S)$). The latter problem can be handled using techniques from the online learning theory, which we discuss in the following section. 
\section{An Online Caching Policy that is Competitive against all Finite-State Prefetchers} \label{online-section}
As the first step towards designing a universal caching policy, we propose a basic online prefetcher that is competitive against the optimal offline single-state (\emph{i.e.,} zeroth-order Markov) prefetcher, where the action of the comparator remains fixed throughout. Subsequently, we show how to extend the proposed prefetcher to compete against multi-state FSPs. We use the classic \emph{Prediction with Expert advice} framework \cite{cesa2006prediction} to design our basic online prefetching policy. This is in sharp contrast with the paper \cite{krishnan1998optimal}, which proposes a problem-specific basic prefetching policy and carries out its analysis using an involved combinatorial method.   
\subsection{Prediction with Expert advice and Online Caching} \label{pred}
For the sake of completeness, we first briefly review the framework of \emph{Prediction with Expert Advice}. Assume that there is a set of $M$ experts. Consider a two-player sequential game played between the learner and an adversary described next. At each round $t$, the adversary selects a reward value $r_{ti} \in [0,1]$ for each expert $i \in [M].$ At the same time (without knowing the rewards for the current round), the learner samples an expert randomly according to a probability distribution $\bm{p}_t$ and accrues the expected reward $\langle \bm{p}_t, \bm{r}_t \rangle.$ The objective of the learner is to achieve a small regret \eqref{regret-def} with respect to the best expert in hindsight. Many variants of the above problem have been studied in the literature and multiple different online policies achieving sublinear regret for this problem are known \cite{vovk1998game, freund1997decision}. 

One of the most fundamental algorithms for the experts problem is \textsc{Hedge} (also known as \textsc{Exponential Weights}). Let the vector $\bm{R}_{t-1} = \sum_{\tau=1}^{t-1}\bm{r}_t$ denote the cumulative rewards of all experts up to round $t-1$. At round $t,$ the \textsc{Hedge} policy chooses the distribution $p^{\text{Hedge}}_{t,i} \propto \exp(\eta R_{t-1})$ for some fixed learning rate $\eta >0.$ It is well-known that the \textsc{Hedge} policy achieves the following regret bound \cite{vovk1998game}:
\begin{eqnarray} \label{reg-bd}
	\max_{i \in [M]} R_{T,i}- \sum_{t=1}^{T} \langle \bm{p}^{\text{Hedge}}_t, \bm{r}_t \rangle \leq \frac{\ln M}{\eta} + \eta \sum_{t=1}^T \sum_{i=1}^M p_{t,i}l_{t,i}^2,
\end{eqnarray}
where $l_{t,i}=1-r_{t,i}$ is the loss of the $i$\textsuperscript{th} expert at round $t$.
\paragraph*{Connection to the Caching problem}
The problem of designing an online prefetcher that competes against a static benchmark can be straightforwardly reduced to the previous experts framework. For this purpose, define an instance of the experts problem with $M=\binom{N}{C}$ experts, each corresponding to a subset of $C$ files. Let the reward $r_{t,i}$ accrued by the $i$\textsuperscript{th} subset at round $t$ be equal to $1$ if the $i$\textsuperscript{th} expert (which corresponds to a particular subset of $C$ files) contains the file $x_t$ requested at round $t$. Else, the value of $r_{t,i}$ is set to zero. A simple but computationally inefficient online caching policy can be obtained by using the \textsc{Hedge} policy on the experts problem defined above. However, a major issue with this naive reduction is that, apparently, it needs to maintain an exponentially large  probability vector $\bm{p}_t$ (with $\binom{N}{C}$ components) at every round $t$, which is clearly computationally infeasible. In a recent paper, we proposed the \textsc{Sage} policy, which gives a near-linear time implementation of the \textsc{Hedge} policy in this context \cite[Algorithm 3]{k-experts}. We now review the \textsc{Sage} policy and show how it can be used in the context of Universal Caching. 
\subsection{The \textsc{Sage} Framework for Online Caching \cite{k-experts}} 
 The \textsc{Sage} framework, proposed in \cite[Algorithm 1]{k-experts}, gives a generic meta-policy that yields an efficient implementation of the \textsc{Hedge} policy by using randomized sampling and exploiting the linearity of the reward function. In particular, we observe that in the online caching problem, the reward accrued by the learner at any round depends only on the \emph{marginal} inclusion probabilities of each file, and not on their joint distribution. Hence, any online learning policy, that yields the same marginal inclusion probabilities as the \textsc{Hedge} policy, achieves the same regret as the \textsc{Hedge} policy. It is inconsequential whether the joint inclusion probabilities are the same or different for these two policies. 
    Based on the above simple observation, the \textsc{Sage} meta-policy works as follows. (a) First, it efficiently computes the marginal file inclusion probabilities induced by the \textsc{Hedge} policy by exploiting the linearity of the reward function. (b) Then it efficiently samples a subset of $C$ files without replacement consistent with the marginals computed in the previous step. In the following, we outline how the above two steps can be carried out efficiently.

   \paragraph{Efficient computation of the  marginal inclusion probabilities} Let the expert $S$ correspond to the subset $S$ of files (with $|S|=C$). The \textsc{Hedge} policy assigns the following probability mass to the expert $S$ at round $t$:
    \begin{equation}
        p_t(S)=\frac{w_{t-1}(S)}{\sum_{S'\subseteq [N]:|S'|=C}w_{t-1}(S')},
    \end{equation}
    where $w_{t-1}(S)\equiv \exp(\eta R_{t-1}(S))$, s.t. $R_{t-1}(S) \equiv \sum_{\tau=1}^{t-1} \mathds{1}(x_\tau \in S)$ denotes the cumulative (offline) cache hits  accrued by the subset $S$ up to round $t,$ and $\eta$ is the learning rate. Consequently, the marginal inclusion probability for the $i$\textsuperscript{th} file is given by:
    \begin{equation}
        \label{eq:marginal}
        p_t(i) = \frac{w_{t-1}(i) \sum_{S \subseteq[N] \backslash\{i\}:|S|=C-1} w_{t-1}(S)}{\sum_{S^{\prime} \subseteq[N]:\left|S^{\prime}\right|=C} w_{t-1}\left(S^{\prime}\right)}.
    \end{equation}
    In the above, we have defined $w_{t-1}(i)\equiv \exp(\eta R_{t-1}(i)),$ where $R_{t-1}(i) \equiv \sum_{\tau=1}^{t-1} \mathds{1}(x_\tau =i)$ denotes the total number of times the $i$\textsuperscript{th} file was requested up to time $t-1$. Both the numerator and the denominator in the probability expression \eqref{eq:marginal} have exponentially many terms and are non-trivial to compute directly. A key observation made in \cite{k-experts} is that both the numerator and denominator  can be expressed in terms of certain \emph{elementary symmetric polynomials} (ESP), which can be efficiently evaluated. To see this, define the vectors $\bfw_t=(w_t(i))_{i\in[N]}$ and $\bfw_{-i,t}=(w_t(i))_{i\in[N]\backslash \{i\}}$. Let $\bfe_k(\cdot)$ denote the ESP of order $k$, defined as follows:
    \[ e_k(\bm{w}) = \sum_{I \subseteq [N], |I|=k} \prod_{j \in I}w_j.\]
With the above definitions in place, the probability term given in \eqref{eq:marginal} can be expressed in terms of ESPs as: 
    \begin{equation}
        \label{eq:esp}
        p_t(i) = \frac{w_t(i)\bfe_{C-1}(\bfw_{-i,t})}{\bfe_C(\bfw_t)}.
    \end{equation}
    
 It is known that any ESP of order $k$ with $N$ variables can be computed efficiently in $\tilde{O}(N)$ time using FFT-based polynomial multiplication methods \cite{shpilka2001lower, grolmusz2003computing}. 
       \paragraph{Sampling without replacement according to a prescribed set of inclusion probabilities} Consider the problem of efficiently sampling a subset of $C$ items without replacement from a universe of $N$ items, where the $i$\textsuperscript{th} item is included in the sampled subset with a prescribed probability $p_i\in[0,1], 1\leq i \leq N$. 
    In other words, if the set $S$ is sampled with probability $\bbP(S)$, then it is required that $\sum_{S: i \in S,|S|=k} \mathbb{P}(S)=p_{i}, \forall i \in[N]$.  
   Given that the inclusion probabilities satisfy the necessary and sufficient condition $\sum_{i=1}^Np_i=k,$ the sampling problem can be efficiently solved using Madow's systematic sampling procedure given below \cite{madow1949theory}. 
    
    \begin{algorithm}
    \caption{Madow's sampling}\label{alg:madow}
    \textbf{Input: } Set $[N]$, size of the sampled set $C$, probability $\bfp$.
    \textbf{Output: } A random set  $S$ containing $C$ elements s.t. $\mathbb{P}(i \in S)=p_i, \forall i.$
        \begin{algorithmic}[1]
            \STATE Let $P_0=0$ and $P_i=P_{i-1}+p_i, \forall i\in[N]$
            \STATE Sample a uniform random variable $U\in[0,1]$.
            \STATE $S \gets \emptyset$
            \FOR{$i\gets 0 \ \text{to} \ k-1$}
                \STATE Select element $j$ if $P_{j-1}\leq U+i\leq P_j$
                \STATE $S\gets S \cup \{j\}$
            \ENDFOR
            \STATE \textbf{return} $S$
        \end{algorithmic}
    \end{algorithm}
    
   Combining part (a) and (b), the overall \textsc{Sage} caching policy is summarized in Algorithm \ref{alg:univcaching}. 
 \begin{algorithm}
    \caption{Online Caching with the \textsc{Sage} framework} \label{alg:univcaching}
    \textbf{Input:} $w(i)=1, \forall i\in [N]$, learning rate $\eta >0.$
    
     \textbf{Output:} A subset of $C$ cached files at every round
    \begin{algorithmic}[1]
        \FOR{$t=1,2,\ldots T$}
            \STATE $w(i)\gets \exp(\eta \mathds{1} \{x_{t-1}=i\}) w(i), \forall i\in[N].$
            \STATE $p(i) \gets \frac{w(i)\bfe_{C-1}(\bfw_{-i})}{\bfe_C(\bfw)}, \forall i \in [N].$ \hfill $\triangleright$ \emph{\textcolor{blue}{Efficient evaluation using FFT}}
            \STATE Sample a set of $C$ files,  with marginal inclusion probabilities $\bfp$ computed as above, using Madow's sampling (Algorithm \ref{alg:madow}).
        \ENDFOR
    \end{algorithmic}
    \end{algorithm}

\subsubsection*{Static regret bound for the \textsc{Sage} policy}
Recall that the quantity $\tilde{\pi}_1(x^T)$ denotes the hit rate achieved by the optimal offline FSP containing a single state. By tuning the learning rate $\eta$ adaptively, the \textsc{Hedge} policy achieves the following data-dependent regret bound \cite[Eqn.\ (14)]{k-experts}:    
\begin{eqnarray} \label{regret-guarantee}
 T(\tilde{\pi}_1-\pi^{\textsc{Hedge}})\leq\sqrt{2Cl_{T}^{*} \ln (N e / C)}+C \ln (N e / C),
   \end{eqnarray} 
where $l_T^* \equiv T- T\tilde{\pi}_1(x^T)$ is the cumulative number of cache misses incurred by the optimal offline caching configuration in hindsight. From the above discussion, it is clear that the \textsc{Sage} policy also achieves the regret bound \eqref{regret-guarantee}. Since $l_T^* \leq T,$ Eqn.\ \eqref{regret-guarantee} trivially yields a sublinear $O(\sqrt{T})$ static regret bound for the online caching problem. Hence, our regret bound \eqref{regret-guarantee} improves upon the previous $O(NC^2\log T \sqrt{T})$ regret bound of the prefetcher (referred to as $``P_1"$) proposed by \cite[Theorem 1, Lemma 3]{krishnan1998optimal}. More importantly, Eqn.\ \eqref{regret-guarantee} gives what is known as a ``small-loss'' bound \cite{lykouris2018small}. In particular, for any request sequence for which the optimal static offline policy concedes a small number of cache-misses (\emph{i.e.,} $l_T^* \ll T$), Eqn.\ \eqref{regret-guarantee} provides a much tighter bound. We will exploit the small-loss bound in our subsequent analysis.

\subsection{Augmenting the \textsc{Sage} policy with a Markovian Prefetcher} 
Equation \eqref{regret-guarantee} gives an upper-bound on the static regret for the \textsc{Sage} caching policy against all offline static prefetchers where the action of the benchmark policy does not change with time. Now consider any given FSM $\mathcal{M}$ containing $S$ number of states. For each state $s \in \mathcal{S},$ let $\bm{x}_s$ be the subsequence of the original file requests obtained by aggregating the requests when the FSM $\mathcal{M}$ was visiting the state $s$. Upon running a separate copy of the \textsc{Sage} policy for each state of the given FSM $\mathcal{M}$, 
we obtain the following regret bound:
\begin{eqnarray}\label{reg-s}
\mathcal{R}_T&=&T(\tilde{\pi}_S^{\mathcal{M}}(x^T) - \pi_S^{\textsc{Sage}}(x^T))\nonumber \\
&=& T\sum_{s=1}^S \big(\tilde{\pi}^{\mathcal{M}}_1 (\bm{x}_s) - \pi_1^{\textsc{Sage}}(\bm{x}_s) \big)\nonumber \\
&\stackrel{(a)}{\leq} & \sum_{s=1}^S \sqrt{2Cl_{T,s}^{*} \ln (N e / C)}+CS \ln (N e / C) \nonumber \\
&\stackrel{(b)}{\leq} & \sqrt{2CSL_{T,S}^*\ln(Ne/C)} + CS \ln (N e / C),
\end{eqnarray}
where $l^*_{T,s}$ denotes the total number of cache misses in the state $s$ incurred by the optimal offline single-state prefetcher and $L^*_{T,S} \equiv \sum_{s=1}^S l^*_{T,s}.$ 
In the above, inequality (a) follows from Eqn.\ \eqref{regret-guarantee} applied to each of the $S$ states of the FSM $\mathcal{M}$ separately, and inequality (b) follows from an application of Jensen's inequality. Specializing the bound \eqref{reg-s} to a $k$\textsuperscript{th} order Markov-prefetcher containing $S=N^k$ many states, we obtain 
\begin{eqnarray} \label{online-regret}
T(\tilde{\mu}_k (x^T)	- \pi_k^{\textsc{Sage}}(x^T)) \leq  \sqrt{2 N^{k}CL_{T,k}^*\ln\frac{Ne}{C}}\nonumber \\
+ N^kC \ln \frac{N e}{C},
\end{eqnarray}
where $L^*_{T,k}$ denotes the minimum number of cache misses incurred by the optimal $k$\textsuperscript{th} order Markovian prefetcher for the file request sequence $x^T.$ Note that the cumulative cache misses $L^*_{T,k}$ could be much smaller than the horizon-length $T$ for many ``regular'' request sequences. Hence, Theorem \ref{online-prefetching-bd} gives a new and tighter adaptive regret bound compared to the previously-known weaker $O(\sqrt{T})$ bound given by \cite[Eqn.\ (24)]{feder1992universal}. 

\textsc{Example 1:} Consider a ``regular'' request sequence $x^T$ generated by an $m$\textsuperscript{th}-order Markovian FSM. By taking $k \geq m,$ we can ensure that $L^*_{T,k} (x^T)=0$ for this sequence. Hence, in this case, the first term on the RHS of the bound \eqref{online-regret} vanishes, resulting in $O(1)$ regret. 

 Combining the regret bound \eqref{reg-s} with Theorem \ref{thm:FSP-MP1}, we have the following guarantee against any FSM containing $S$ many states:
\begin{theorem} \label{online-prefetching-bd}
	For any file request sequence $\bm{x}^T,$ the regret of the $k$\textsuperscript{th} order Markovian FSM running a separate copy of the \textsc{Sage} caching policy on each state, compared to an optimal offline FSP containing at most $S$  states, is upper-bounded as: 
	\begin{eqnarray*} \label{reg-bd-expr}
 \mathcal{R}_T\equiv T(\tilde{\pi}_S(x^T) - \pi_k^{\textsc{Sage}}(x^T)) \leq  	T\min \bigg(1-C/N,	\nonumber \\\sqrt{ \frac{{\ln S}}{2(k+1)}}\bigg) 
+\sqrt{2N^{k}CL_{T,k}^*\ln\frac{Ne}{C}} +
	 N^kC \ln \frac{N e}{C}. \hspace{6pt}
	\end{eqnarray*}
\end{theorem}
\textsc{Example 2:} Consider a file request sequence $\bm{x}_Q^T$ generated by any FSM containing at most $Q$ states. The FSM needs not be Markovian (c.f. \textsc{Example 1}). Refer to Section \ref{gen-sec} of the Appendix for details on the request sequence generation. Combining Theorem \ref{thm:FSP-MP1} and Theorem \ref{online-prefetching-bd}, we have: 
\begin{eqnarray} \label{cache-miss-hedge1}
&&\text{Expected fraction of cache misses conceded by the} \nonumber \\ 
&& \text{\textsc{Sage} policy used 
with a $k$\textsuperscript{th} order Markovian FSM }  \nonumber \\  
&\leq & \min \bigg(1-\frac{C}{N},\sqrt{\frac{\ln Q}{2(k+1)}}\bigg) + \nonumber \\ &&\sqrt{\frac{2N^{k}C}{T} \ln \frac{Ne}{C}\min\bigg(1-\frac{C}{N}, \sqrt{\frac{\ln Q}{2(k+1)}}}\bigg) \nonumber \\
&&+ \frac{N^k C}{T} \ln \frac{Ne}{C}, 
\end{eqnarray}
where we have used the fact that an optimal FSM with $Q$ many states incurs \emph{zero} cache misses for the request sequence $\bm{x}_Q^T$. If the value of $Q$ is known (however, the structure of the FSM remains unknown), the optimal order $k^*$ of the Markovian prefetcher minimizing the upper bound in Eqn.\ \eqref{cache-miss-hedge1} can be computed using calculus. From Eqn.\ \eqref{cache-miss-hedge1}, it also follows that for any fixed value of $Q$, the expected fraction of cache-misses can be made approach to zero at the rate of $O(T^{-1/2})$ by taking $k \gg \ln Q$. See Section \ref{expts} for the numerical results.

In the following section, we design a universal caching policy that achieves a sublinear $O\big((\log T)^{-1/2}\big)$ regret bound  for all file request sequences against any FSP containing \emph{unknown and arbitrarily many states} $S$. 
 
 \section{A Universal Caching Policy} \label{universal-caching-section}
  In Theorem \eqref{online-prefetching-bd}, we are free to choose the order $k$ of the Markovian FSM as a function of the (known) horizon-length $T$. Since the number of states $S$ in the benchmark comparator could be arbitrarily large, it is clear that in order to achieve asymptotically zero regret (normalized w.r.t. the horizon length $T$), the order $k$ of the Markovian prefetcher needs to be increased accordingly with $T$. By setting $N^k = O(\frac{T}{\log T})$ in Theorem \ref{online-prefetching-bd}, we obtain the following bound on the regret against \emph{all} FSPs having arbitrarily many states: 
 $ \frac{\mathcal{R}_T}{T} \leq 	O(\frac{1}{\sqrt{\log T}}).$
\subsection*{Efficient implementation using Lempel-Ziv (LZ) parsing}
Similar to the binary prediction problem considered in \cite{feder1992universal}, we can use incremental parsing algorithms, such as Lempel-Ziv'78 \cite{ziv1978compression} to adaptively increment the order of the Markovian Prefetcher when the horizon length $T$ is not known a priori \cite{krishnan1998optimal, vitter1996optimal}. However, instead of constructing a binary parse tree as in \cite{feder1992universal}, we build an $N$-ary tree with the \textsc{Sage} policy running at each node. 
In particular, the LZ parsing algorithm parses the $N$-ary request sequence into distinct phrases such that each phrase is the shortest phrase that is not a previously parsed phrase. In the parse tree, each new phrase corresponds to a leaf in the tree. The parsing proceeds as follows: the LZ tree is initialized with a root node and $N$ leaves. The current tree is used to create the next phrase by following the path from the root to leave according to the consecutive file requests. Once a leaf node is reached, the tree is extended by making the leaf an internal node, and adding $N$ offsprings to the tree and then moving to the root of the tree. Each node of the tree now corresponds to a state of the Markovian prefetcher and runs a separate instance of the \textsc{Sage} policy.  A classical result, established in  \cite[Theorem 2]{lempel1976complexity}, tells that the number of nodes in an $N$-ary LZ tree generated by an arbitrary sequence of length $T$ grows sub-linearly as $c(T)= O(\frac{T \log N}{\log T}).$ Hence, for any fixed $k$, the fraction of file requests made on a node with depth less than $k$ vanishes asymptotically. Hence, the expected fraction of cache hits $\pi^{\textsc{LZ}}$ achieved by the LZ prefetcher is asymptotically lower bounded by that of a $k$\textsuperscript{th} order Markovian FSP containing $N^k \approx c(T)$ states up to a sublinear regret term. The following theorem makes this statement precise.


\begin{theorem}\label{thm:IP-MP}
    For any integer $k\geq 0$, the regret of the \textsc{LZ} prefetcher w.r.t. an offline $k$\textsuperscript{th} order Markovian prefetcher can be upper-bounded as:
    $$\mathcal{R}_T \equiv T(\tilde{\mu}_k - \pi^{\textsc{LZ}}) \leq   \delta(c(T),L_T^{*,LZ})+ kc(T),$$ where $c(T) \equiv O(\frac{T \log N}{\log T})$ and $\delta(B,l_T^*) \equiv \sqrt{2BCL_{T}^{*,LZ} \ln (N e / C)}+CB \ln (N e / C)$.
\end{theorem}
See Section \ref{proof:ip-mp} of the Appendix for the proof.





\section{Conclusion}\label{conclusion}
In this paper, we proposed an efficient online universal caching policy that results in a  sublinear regret against all finite-state prefetchers containing arbitrarily many states. We presented the first data-dependent regret bound for the universal caching problem by making use of the \textsc{Sage} framework \cite{k-experts}. In the future, it will be interesting to extend these techniques to other online learning problems to design policies with improved regret guarantees. Furthermore, designing universal algorithms that also guarantee sublinear \emph{dynamic} regret and take into account the file download costs will be of interest. 
\section{Acknowledgment} \label{ack}
This work was partly supported by a grant from the DST-NSF India-US collaborative research initiative under the TIH at the Indian Statistical Institute at Kolkata, India.
\clearpage
\balance
\bibliographystyle{IEEEtran}
\bibliography{references}
\clearpage
\nobalance
\twocolumn[\section*{\Large Appendix} \vspace{1cm}]
\section{Proofs}
\label{sec:proofs}


%
%
\subsection{Proof of Theorem \ref{thm:FSP-MP1}}\label{proof:fsp-mp}
%
%
Consider an FSP $\mathcal{M},$ containing $S$ states, whose state transition function is given by $g$.  
For any integer $j \geq 1,$ construct a new FSP $\mathcal{\tilde{M}}$ whose state at round $t$ is given by $\tilde{s}_t= (s_t, x^{t-1}_{t-j}).$ In other words, the states of the FSP $\tilde{\mathcal{M}}$ are constructed by juxtaposing the states of $\mathcal{M}$ and the $k$\textsuperscript{th} order Markov prefetcher. The transition function of $\tilde{\mathcal{M}}$ is naturally defined as $\tilde{g}((s_t, x^{t-1}_{t-j}), x_t)= (g(s_t), x^t_{t-j+1}).$ Consequently, if both the FSPs $\mathcal{M}$ and $\mathcal{\tilde{M}}$ are fed with the same file request sequence $\bm{x}$, the current state of the FSP $\mathcal{M}$ can be read off from the first component of the current state of the FSP $\mathcal{\tilde{M}}.$ Let $\tilde{\mu}_{j,S}(x^T)$ be the hit rate achieved by the new FSP $\tilde{\mathcal{M}}$ for the given file request sequence. Define a sequence of random variables $(X_1, X_2, \ldots )$, which are distributed according to the empirical probability measure induced by the consecutive terms of the given file request sequence $x_1^T.$ In other words, for any natural number $k,$ we define the joint distribution:
\begin{eqnarray*}
&&\mathbb{P}\big((X_1, X_2, \ldots, X_k)= (z_1, z_2, \ldots, z_k)\big)\\
&=& \frac{|\{q: (x_q, x_{q+1}, \ldots, x_{q+k-1})= (z_1, z_2, \ldots, z_k) \}|}{T}.	
\end{eqnarray*}

 Recall that the quantity $\tilde{\mu}_j(x^T)$ denotes the cache hit rate achieved by the $k$\textsuperscript{th} order Markovian prefetcher. We first establish the following technical result. 
\begin{framed}
\begin{proposition} \label{basic-prop}
For any file request sequence $x^T,$ we have
\begin{eqnarray*}
&&\tilde{\mu}_{j,S}(x^T) - \tilde{\mu}_j(x^T)\leq \min\bigg(1-C/N, \\
&& \sqrt{\frac{\ln 2}{2}\big(H(X_{j+1}|X^j)-H(X_{j+1}|X^j,S)\big)}\bigg).
\end{eqnarray*}

\end{proposition}
\end{framed}
\begin{proof}
Corresponding to the state $\tilde{s}_t \equiv (s_t, x^{t-1}_{t-j})= (s, x^j)$  of the FSP $\tilde{\mathcal{M}}$, let $\tilde{p}_{s,x^j}(\cdot)$ be the conditional probability distribution for the next file request $x_t.$  For any r.v. $Z$, let the quantity $\mathbb{E}_{Z}(\cdot)$ denote the expectation w.r.t. the empirical distribution of the r.v. $Z$. Since the optimal offline policy caches the most requested $C$ files on any state, we have the following sequence of bounds
\begin{eqnarray} \label{bd1}
&&\tilde{\mu}_{j,S}(x^T) - \tilde{\mu}_j(x^T)	\nonumber \\
&=& \mathbb{E}_{S, X^j} \bigg(\max_{A: |A|=C} \sum_{i\in A} p_{S,X^j}(X_{j+1}=i)- \nonumber\\
&& \max_{A: |A|=C} \sum_{i \in A} p_{X^j}(X_{j+1}=i)\bigg)\nonumber\\
&\stackrel{(a)}{\leq} & \mathbb{E}_{S, X^j}\big( \max_{A: |A|=C} \sum_{i \in A} (p_{S, X^j}(i)- p_{X^j}(i))\big) \nonumber \\
& \leq & \mathbb{E}_{S, X^j}\big(\textsc{TV}(p_{S, X^j}(X_{j+1}), p_{X^j}(X_{j+1})) \big) \nonumber \\
&\stackrel{(b)}{\leq} & \mathbb{E}_{S, X^j} \bigg(\sqrt{\frac{\ln 2}{2}D(P(X_{j+1}|S,X^j)||P(X_{j+1}|X^j))}\bigg) \nonumber \\
&\stackrel{(c)}{\leq}&  \sqrt{\frac{\ln 2}{2}D(P(X_{j+1}|S,X^j)||P(X_{j+1}|X^j))}\nonumber \\
&=& \sqrt{\frac{\ln 2}{2}(H(X_{j+1}|X^j)-H(X_{j+1}|X^j,S)) }.
\end{eqnarray}
where (a) follows from the sub-additivity of the $\max (\cdot)$ function, (b) follows from Pinsker's inequality, and (c) follows from the concavity of the square root function and Jensen's inequality. Furthermore, we also have 
\begin{eqnarray}
&&\tilde{\mu}_{j,S}(x^T) - \tilde{\mu}_j(x^T)\nonumber	\\
&=& \mathbb{E}_{S, X^j} \bigg(\max_{A: |A|=C} \sum_{i\in A} p_{S,X^j}(X_{j+1}=i)- \nonumber\\
&& \max_{A: |A|=C} \sum_{i \in A} p_{X^j}(X_{j+1}=i)\bigg) \nonumber\\
&\leq& 1-C/N.\label{bd2}
\end{eqnarray}
Combining the bounds \eqref{bd1} and \eqref{bd2} completes the proof of Proposition \ref{basic-prop}. 
\end{proof}
We now establish Theorem \ref{thm:FSP-MP1}. 
Since the current state of the FSP $\mathcal{M}$ is a deterministic function of the current state of the FSP $\tilde{\mathcal{M}},$ it immediately follows that $\tilde{\pi}_S(x^T) \leq \tilde{\mu}_{j,S} (x^T)$ for any $j\geq 1.$ By the same argument, we have $\tilde{\mu}_{k}(x^T) \geq \tilde{\mu}_j(x^T), \forall k\geq j.$ Hence, we have 
%
    \begin{eqnarray*}
        &&\tilde{\pi}_{S}(x^T) -\tilde{\mu}_k(x^T)\\ 
        &\leq & \tilde{\mu}_{j,S}(x^T) -\tilde{\mu}_k(x^T) \\
        &\leq& \frac{1}{k+1} \sum_{j=0}^k [\tilde{\mu}_{j,S}(x^T)-\tilde{\mu}_j(x^T)]\\
        &\stackrel{(a)}{\leq}& \frac{1}{k+1} \sum_{j=0}^k \min\bigg(1-C/N,\\
        &&\sqrt{{\frac{\ln2}{2}} (H(X_{j+1}|X^j)-H(X_{j+1}|S,X^j))}\bigg)\\
        &\stackrel{(b)}{\leq}& \min\bigg(1-C/N, \\
        &&  \sqrt{ \frac{{\ln2}}{2(k+1)}\big(\sum_{j=0}^k H(X_{j+1}|X^j)-H(X_{j+1}|S,X^j)\big)}\bigg)\\
        &\stackrel{(c)}{=}& \min \bigg(1-C/N, \\ &&\sqrt{ \frac{{\ln2}}{2(k+1)}\big(H(X^{k+1})- H(X^{k+1}|S)\big)}\bigg) \\
        &=& \min\bigg(1-C/N, \sqrt{ \frac{{\ln2}}{2(k+1)}I(S;X^{k+1})}\bigg) \\
        &\stackrel{(d)}{\leq} & \min\bigg(1-C/N, \sqrt{ \frac{{\ln S }}{2(k+1)}}\bigg), 
    \end{eqnarray*}
    where in (a), we have used Proposition \ref{basic-prop} and in (b), we have Jensen's inequality twice (using the concavity of the $\min(\cdot)$ and the $\sqrt(\cdot)$ functions), in (c), we have used the chain rule for entropy, and in (d), we have trivially upper bounded the mutual information by $\log S $.
    $\hfill \blacksquare$
%



\subsection{Proof of Theorem \ref{thm:IP-MP}}
\label{proof:ip-mp}
\begin{proof}
Let $c(T)$ denote the total number of nodes in the LZ tree after time $T$ and $L_T^{*,LZ}$ denote the number of cache misses by the optimal offline prefetching policy which follows the same tree growth process as the online LZ parsing algorithm (but could prefetch files different from that of the online policy).
 Using the bound \eqref{reg-s} on each node of the LZ parse tree and applying Jensen's inequality, we get 
     \begin{equation}\label{eq:bound-using-jensen}
        T\pi^{\textsc{LZ}} + \delta(c(T),L_T^{*,LZ})\geq\sum_{s=1}^{c(T)} \max_{\bfy\in\calY} \inner{\bfy,\bfX_s(T)},
    \end{equation}
    where $\bfX_s(T)$ denotes the aggregate count vector of all requests made up to the time $T$ while the LZ tree was visiting the state $s$. We now lower bound the RHS of the above inequality in terms of the total hits achieved by a $k$\textsuperscript{th}-order Markovian prefetcher. For any fixed $k\geq 0,$ let $J_1$ be the set of states labeled by strings of length less than $k$ and $J_2$ is the remaining set of states in the LZ tree. We can decompose the total cache hits as follows:    
    \begin{eqnarray}\label{eq:split}
        &&\sum_{s=1}^{c(T)} \max_{\bfy\in\calY} \inner{\bfy,\bfX_s(T)}\nonumber \\
        &=&\sum_{s\in J_1} \max_{\bfy\in\calY} \inner{\bfy,\bfX_s(T)}+\sum_{s\in J_2} \max_{\bfy\in\calY} \inner{\bfy,\bfX_s(T)} \nonumber \\
        &\geq& \sum_{s\in J_2} \max_{\bfy\in\calY} \inner{\bfy,\bfX_s(T)}.
    \end{eqnarray}
    
    
    The states in $J_2$ form a refinement of an order-$k$ Markovian prefetcher, since each state in $J_2$ is labeled by strings of length at least $k$ \cite{feder1992universal}. Hence, the second term in \eqref{eq:split} can be lower bounded by the number of cache hits accrued by the order-$k$ Markovian prefetcher for the requests made on the states in $J_2$. Since the total number of parsed strings is $c(T)$ and only the first $k$ requests of any parsed string are included in $J_1,$ the total number of requests made on the bins in $J_2$ is at least $T-kc(T).$ Hence, the quantity \eqref{eq:split} can be further lower bounded by $(T-kc(T))\hat{\mu}_k \geq T\hat{\mu}_k - kc(T),$ where $T\hat{\mu}_k$ is the total number of cache-hits by the order-$k$ Markovian prefetcher over the entire input sequence. Substituting the lower bounds in Eqn.\ \eqref{eq:bound-using-jensen} we get
    \begin{equation*}
        T\pi^{LZ} + \delta(c(T),L_T^*)\geq  T\hat{\mu}_k-kc(T).
    \end{equation*}
    Rearranging the above, we get the final bound
     \begin{equation*}
        \mathcal{R}_T= T(\hat{\mu}_k-\pi^{LZ}) \leq \delta(c(T),L_T^*) + kc(T).
    \end{equation*}
    \end{proof}
    
  \subsection{\textsc{LRU} and \textsc{FIFO} are Finite State Prefetchers} \label{fsp-examples}  
  In this section, we argue that \textsc{LRU} and \textsc{FIFO} caching policies belong to the class of the Finite State Prefetchers. To prove the claim, we need to construct FSPs that simulate the \textsc{LRU} and the \textsc{FIFO} policies, respectively.


Recall that LRU is a cache replacement policy that evicts the least-recently requested file from the cache to store the newly requested file that is not in the cache. Hence, the LRU policy caches the $C$ most-recently requested files at all times. 
\paragraph*{\textsc{LRU}} Let each state $\sigma \equiv (\sigma_1,\ldots,\sigma_C)$ correspond to most-recently requested $C$ distinct files ordered in the increasing order of the time-stamps of their latest requests. In other words, the file $\sigma_C$ was requested most recently and the file $\sigma_1$ is the $C$\textsuperscript{th} most-recently requested file. The prefetching function for LRU is defined as
$$f^{\textsc{LRU}}(\sigma)=\sigma.$$
In other words, the prefetcher caches the $C$ most-recently requested files at each state.
Suppose, at the next round, the file $x\in [N]$ is requested. The state-transition function $g$ is defined as: 
\begin{eqnarray*}
	g^{\textsc{LRU}}(\sigma,x)=\begin{cases}(\sigma_1, \sigma_2, \ldots \sigma_{i-1}, \sigma_{i+1}\ldots, \sigma_{C-1},\sigma_C,x),\\
	~~\text{if} \ x \in \{\sigma_{1},\ldots,\sigma_C\} \text{ and } x=\sigma_i, i\in N ,\\
	(\sigma_2, \sigma_3, \ldots, \sigma_{C-1},\sigma_C,x), ~ \textrm{otherwise}.	
\end{cases}
\end{eqnarray*}

 It can be seen that the state transition function maintains the correct ordering of the files according to their latest requests at all times. The starting state $s_0$ can be selected arbitrarily.  The total number of states in this construction is $|\calS^{\textsc{LRU}}|=N(N-1)\ldots (N-C+1) \leq N^C.$ This completes our description of \textsc{LRU} as an FSP. 

\paragraph*{\textsc{FIFO}} Similar to the above construction, we now show that the \textsc{FIFO} caching policy can also be simulated by an FSP. Recall that \textsc{FIFO} is a cache replacement policy which, while inserting a newly requested file that is not in the cache, evicts the oldest file from the cache. 

Let the state $\sigma \equiv (\sigma_1,\ldots,\sigma_C)$ correspond to the most-recently cached $C$ distinct files ordered in the increasing order of the time-stamps of their insertion to the cache under the \textsc{FIFO} policy. In other words, the file $\sigma_1$ is the oldest file in the cache and the file $\sigma_C$ is the newest file in the cache. The prefetching function is simply given by $$f^{\textsc{FIFO}}(\sigma)=\sigma.$$
Suppose that at time $t$, the file $x\in [N]$ was requested. The state transition function is given by: 
\begin{equation*}
    g^{\textsc{FIFO}}(\sigma,x)= 
    \begin{cases}
    \sigma, \ \text{if} \ x \in \{\sigma_{1},\ldots,\sigma_C\}, \\
        (\sigma_2, \sigma_3, \ldots, \sigma_{C-1},\sigma_C,x), \ \text{otherwise}. 
    \end{cases}
\end{equation*}
Note that if the newly requested file is already in the cache, the state of the FSP does not change. On the other hand, when a non-cached file is requested, it is placed at the end of the state and the entire tuple is shifted to the left by one place and the file $\sigma_1$ is evicted. This is precisely the FIFO policy. The total number of states in this construction is $N(N-1)\ldots (N-C+1) \leq N^C.$

    \section{Experiments} \label{expts}


In this section, we report some simulation results demonstrating the practical efficacy of the proposed universal caching policy \footnote{Code available at \url{https://github.com/AtivJoshi/UniversalCaching}}. In our simulations, we use a synthetic file request sequence generated by a randomly constructed Finite State Machine. The structure and the number of states of the FSM remain hidden from the online policies that we evaluate. The details of the setup and the simulation results are discussed below. 


\begin{algorithm}
\caption{Synthetic File Request Generation} \label{data-gen}
\textbf{Input:} Number of states $Q$, cache size $C$, transition function $g$, initial state $s_0$, arrays of files $A_s, |A_s|=C, \forall s \in \mathcal{S},$ horizon length $T$.

\textbf{Output:} Generated file request sequence $\bfx^T=\{x_1,\cdots,x_T\}$.
\begin{algorithmic}[1]
    \STATE Initialize $s\gets s_0$.
    \FOR{$t\gets 1 \ \text{to} \ T$}
        \STATE Pick a file $x_t$ uniformly at random from the set $A_s$.
        \STATE $s\gets g(x_t, s)$
    \ENDFOR
    \STATE \textbf{return} $x^T$
\end{algorithmic}
\end{algorithm}

\subsection{Synthetic Data Generation} \label{gen-sec}
We generate a synthetic file request sequence that can be perfectly predicted by some Finite State Predictor with zero cache misses. One such simple request generation scheme is outlined in Algorithm \ref{data-gen}. In this scheme, we randomly construct an FSM $\mathcal{M}$ containing $Q$ many states. For this, we initialize a random transition function $g(\cdot,\cdot)$ by generating a random $Q \times N$ matrix with entries in $\{1,\cdots,Q\}$. For each state $s \in \mathcal{S}$, we randomly sample an array $A_s$ containing $C$ files. We also randomly select an initial state $s_0$. Once constructed, we use the same FSM $\mathcal{M}$ for generating the entire file request sequence. To generate the request sequence, we start from the initial state $s_0$ and at each state $s_t,$ we randomly pick a file $x_t$ from the set $A_{s_t}$ uniformly at random. Then we go to the next state $s_{t+1}=g(x_t,s_t)$ and repeat the process up to time $T$. 

Note that the FSP characterized by the FSM $\mathcal{M}$ and the prediction function $f(s)=A_s$ predicts the generated request sequence with 100\% accuracy. 
%
\begin{figure}
    \begin{subfigure}{0.49\textwidth}
            \includegraphics[trim={0 0 0 3cm},clip,scale=0.49]{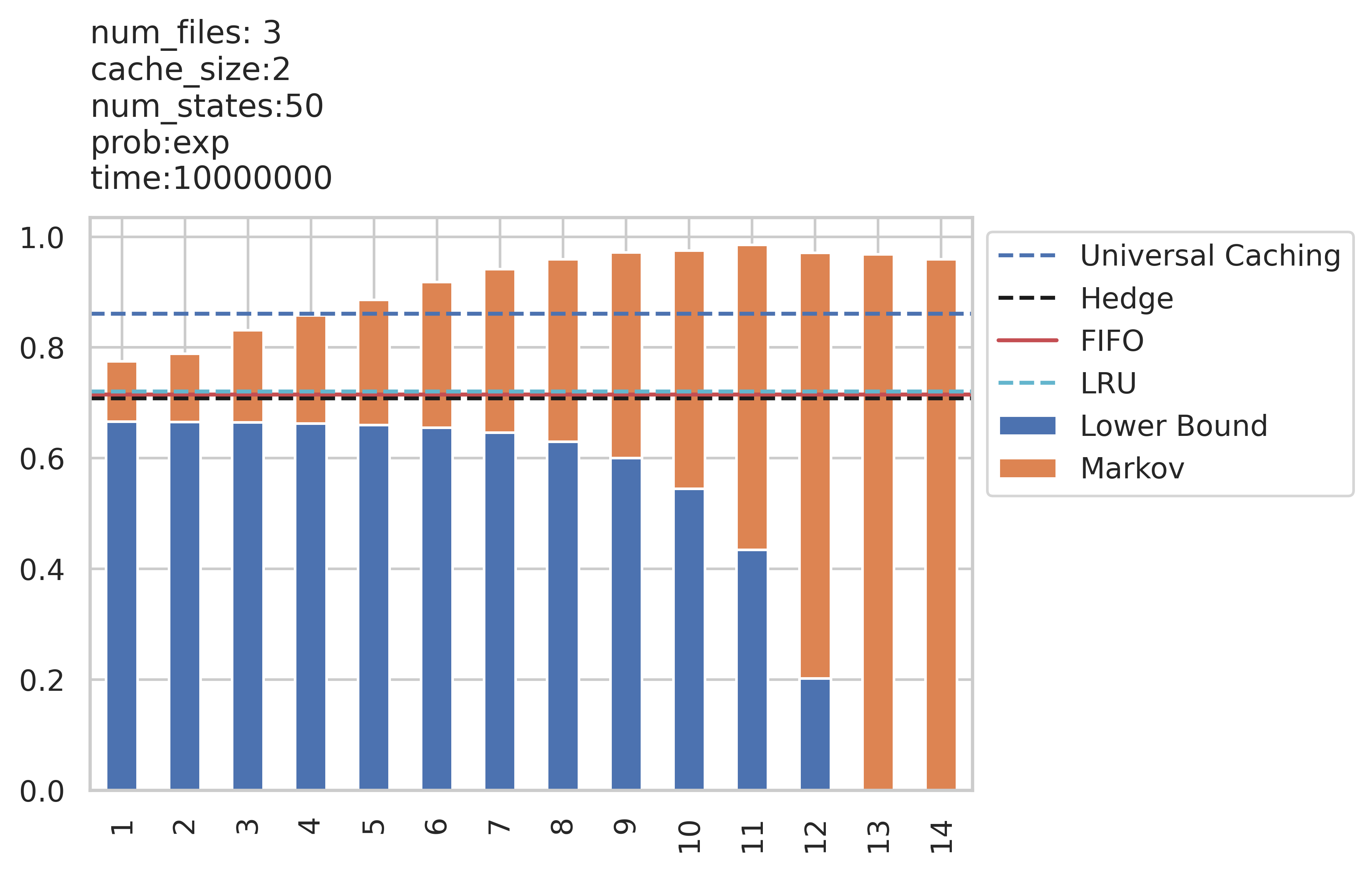}
            \put(-210,-5){\small{Order $k$ of the Markov predictor}}
            \caption{$N=3$, $C=2$, $Q=50$, $T=10M$}
            \label{subfig:four}
    \end{subfigure} \hfill
   \begin{subfigure}[]{0.49\textwidth}
        \includegraphics[trim={0 0 0 3cm},clip,scale=0.48]{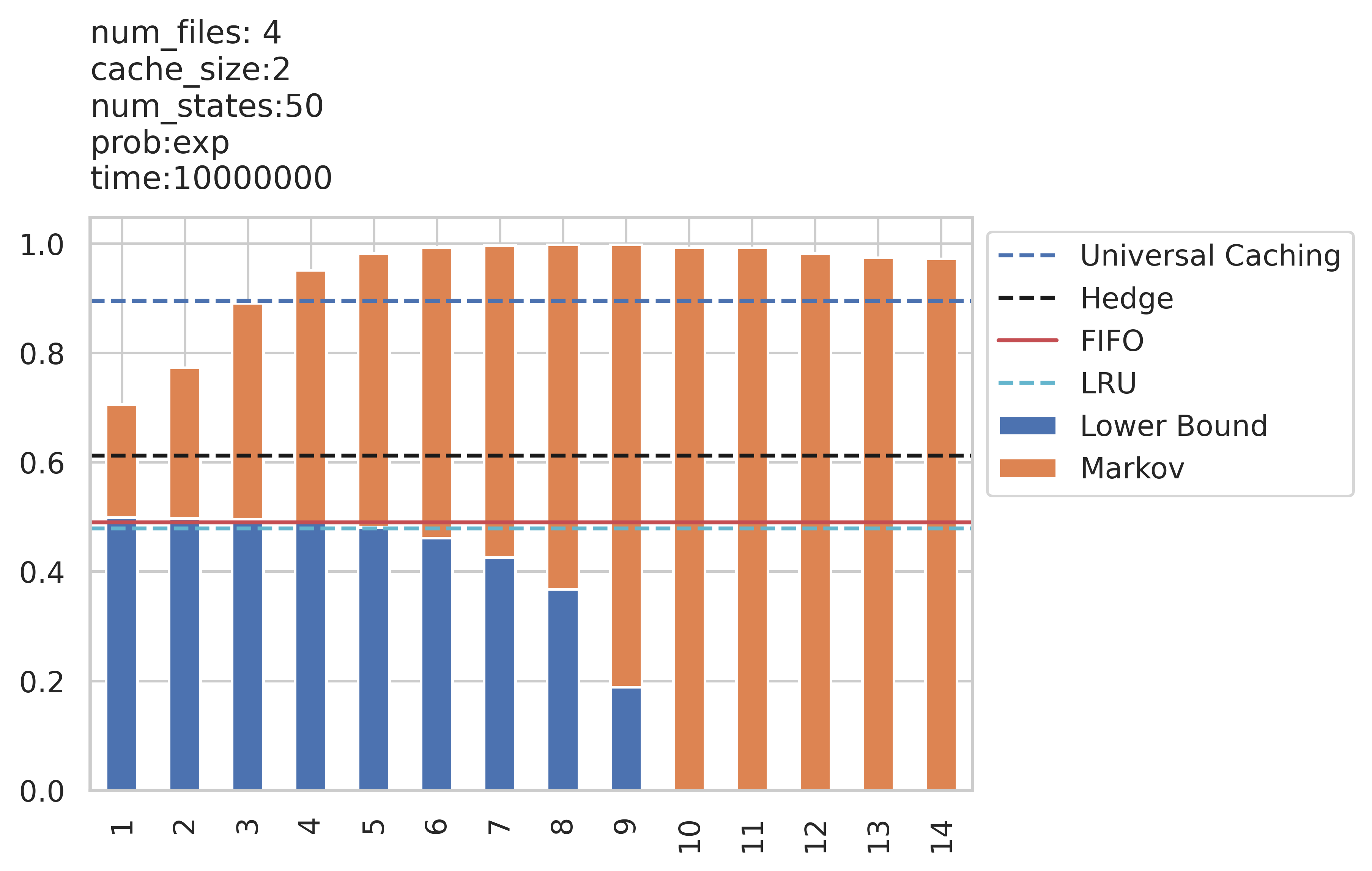}
        \put(-210,-5){\small{Order $k$ of the Markov predictor}}
        \caption{$N=4$, $C=2$, $Q=50$, $T=10M$}
        \label{subfig:two}
    \end{subfigure}
%
    \caption{Plots depicting the lower bounds (given by Eqn.\ \eqref{cache-miss-hedge1}) and the actual hit rate achieved by different policies on the synthetic file request sequence generated a randomly constructed FSM for two different parameter settings. It appears that the bound in Eqn.\ \eqref{cache-miss-hedge1} is quite conservative and the observed performance of the proposed universal policy far exceeds the lower bound. Furthermore, the universal caching policy and the $k$\textsuperscript{th} order FSP for a suitable value of $k$ achieves very high hit rate compared to a vanilla \textsc{Sage} policy. }
    \label{fig:hitrate}
\end{figure}

\subsection{Results and Discussion} 
The numerical simulation results are shown in Figure \ref{fig:hitrate}. The hit rate achieved by the $k$\textsuperscript{th} order FSM is shown by the saffron bars as a function of the order $k$. The blue bars represent the lower bound \eqref{cache-miss-hedge1} on the hit rate of the $k$\textsuperscript{th} order Markov predictors. The horizontal dotted red line denotes the hit rate achieved by the \textsc{Sage} policy. Finally, the horizontal dotted blue line denotes the hit rate achieved by the Universal Caching policy. From the plots, we see that both the universal caching policy and the $k$\textsuperscript{th} order Markovian FSP perform exceptionally well compared to the vanilla \textsc{Sage} policy, which is competitive only against a static benchmark. Furthermore, the corresponding lower bound to the hit rate given by Eqn.\ \eqref{cache-miss-hedge1} seems to be loose compared to the observed performance.

\end{document}